\documentclass[]{raa}            
\usepackage{graphicx,times}
\usepackage{amssymb}
\usepackage{amsmath}
\usepackage{natbib}
\begin{document}
   \title{The preliminary statistical analysis of LAMOST DR8 low resolution AFGK stars
}

\volnopage{ {\bf 0000} Vol.\ {\bf 0} No. {\bf XX}, 000--000}
\setcounter{page}{1}

\author{Y. H. Chen\inst{1,3,4} \ G. W. Li\inst{2} \ H. Shu\inst{1,3,4}}

\institute{\inst{1} Institute of Astrophysics, Chuxiong Normal University, Chuxiong 675000, China; {yanhuichen1987@126.com\\
           \inst{2} National Astronomical Observatories, Chinese Academy of Sciences, Beijing 100101, China\\
           \inst{3} School of Physics and Electronical Science, Chuxiong Normal University, Chuxiong 675000, China\\
           \inst{4} Key Laboratory for the Structure and Evolution of Celestial Objects, Chinese Academy of Sciences, P.O. Box 110, Kunming 650011, China}
\\
\vs \no
{\small Received [0000] [July] [day]; accepted [0000] [month] [day] }}

\abstract{We download the LAMOST DR8 low resolution catalog 6,478,063 AFGK tpye stars and plot the figures of effective temperature, gravitational acceleration, and metal abundance. Some small and medium mass stars are evolved from pre-main sequence or main sequence stage to planetary nebula stage or white dwarf stage by the stellar evolution code \texttt{MESA}. We analyze the observed statistical data and model calculation results, and then obtain some basic conclusions preliminarily. Most red giant and asymptotic giant stars with log$g$ less than 0.85 have poor metal abundance. Most hot A type main-sequence stars are metal rich stars with log$g$ from 3.5 to 4.5. The conclusions are reasonable within a certain error range. The theory of a gap area in the H-R diagram for stellar evolutions of medium mass stars is reflected in the statistical figures. The central core hydrogen burning stage and the central core helium burning stage correspond to the peak structures in the gravitational acceleration statistical figures respectively. The metal abundances among A, F, G, and K type stars have a wide distribution. We can not simply replace the metal abundances of these stars with the metal abundance of the Sun when doing a fine research work.
 }

\authorrunning{Y. H. Chen, G. W. Li, H. Shu}            
\titlerunning{The preliminary analysis of LAMOST DR8 AFGK stars}  
\maketitle

\section{Introduction}

Photometric observation and spectral observation are important methods to detect celestial information. The NASA $Kepler$ space telescope was launched in 2009 and retired in 2018 (Thompson et al. 2018, C$\acute{o}$rsico 2020). The high-precision continuous photometric data of tens of thousands of stars have brought revolutionary development to the research of exoplanets and astrophysics. The observed data of $Kepler$ 2 can be used to study young open clusters, bright stars, galaxies, supernovae, and asteroseismology (Howell et al. 2014). The Sloan Digital Sky Survey ($SDSS$) telescope is designed to measure multi-band and multi-color photometric and spectral data of order a million celestial bodies, and then calculate the red-shift values (Gunn et al. 2006). The $SDSS$ is also used to research the DA type white dwarfs (Tremblay et al. 2011, Gianninas et al. 2011), the magnetic fields of DA type white dwarfs (K$\ddot{u}$lebi et al. 2009), the supernova legacies (Astier et al. 2006), and the near-Earth objects (Raymond et al. 2004).

The Large sky Area Multi-Object fiber Spectroscopic Telescope (LAMOST) (Cui et al. 2012), also named the Guo Shou Jing Telescope (GSJT), is an optical telescope developed by Chinese scientists. LAMOST has a wide field of view ($\sim$5 degrees) and a large effective aperture ($\sim$4 meters) (Cui et al. 2012). LAMOST will observe the spectra of millions of objects in the northern sky and do the research work of the evolution history of galaxies, the distribution of dark matter, the sub-structures in the Milky Way halo, the black hole in the center of the Milky Way, and other cutting-edge topics (Zhao et al. 2012).

Since 2011, LAMOST has released a large number ($\sim$millions) of spectra every year, including galaxies, quasi-stellar objects (QSOs), stars, AFGK type stars, M type stars, A type stars, and unknown objects for the low resolution catalog. We download the low resolution catalog AFGK type stars from the data release 8 (DR8) v1.0 (the corresponding URL http://www.lamost.org/dr8/) and do some basic research work. In Sect. 2, we show a basic analysis of observed data of LAMOST DR8 low resolution catalog AFGK stars. In Sect. 3, we show a preliminary analysis of observation results based on evolutionary models. Then, a discussion and conclusions is given in the last section.

\section{The basic analysis of observed data of LAMOST DR8 low resolution catalog AFGK stars}

The LAMOST DR8 v1.0 contains 219,776 galaxies, 71,786 QSOs, 6,478,063 AFGK tpye stars, and so on from 2011 to 2020. We download the 6,478,063 AFGK tpye stars for the low resolution catalog and plot the figures of effective temperature ($T_{eff}$), gravitational acceleration (log$g$), and metal abundance ([Fe/H]). Then, we do a basic analysis based on the figures.

\begin{figure}
\begin{center}
\includegraphics[width=10.5cm,angle=0]{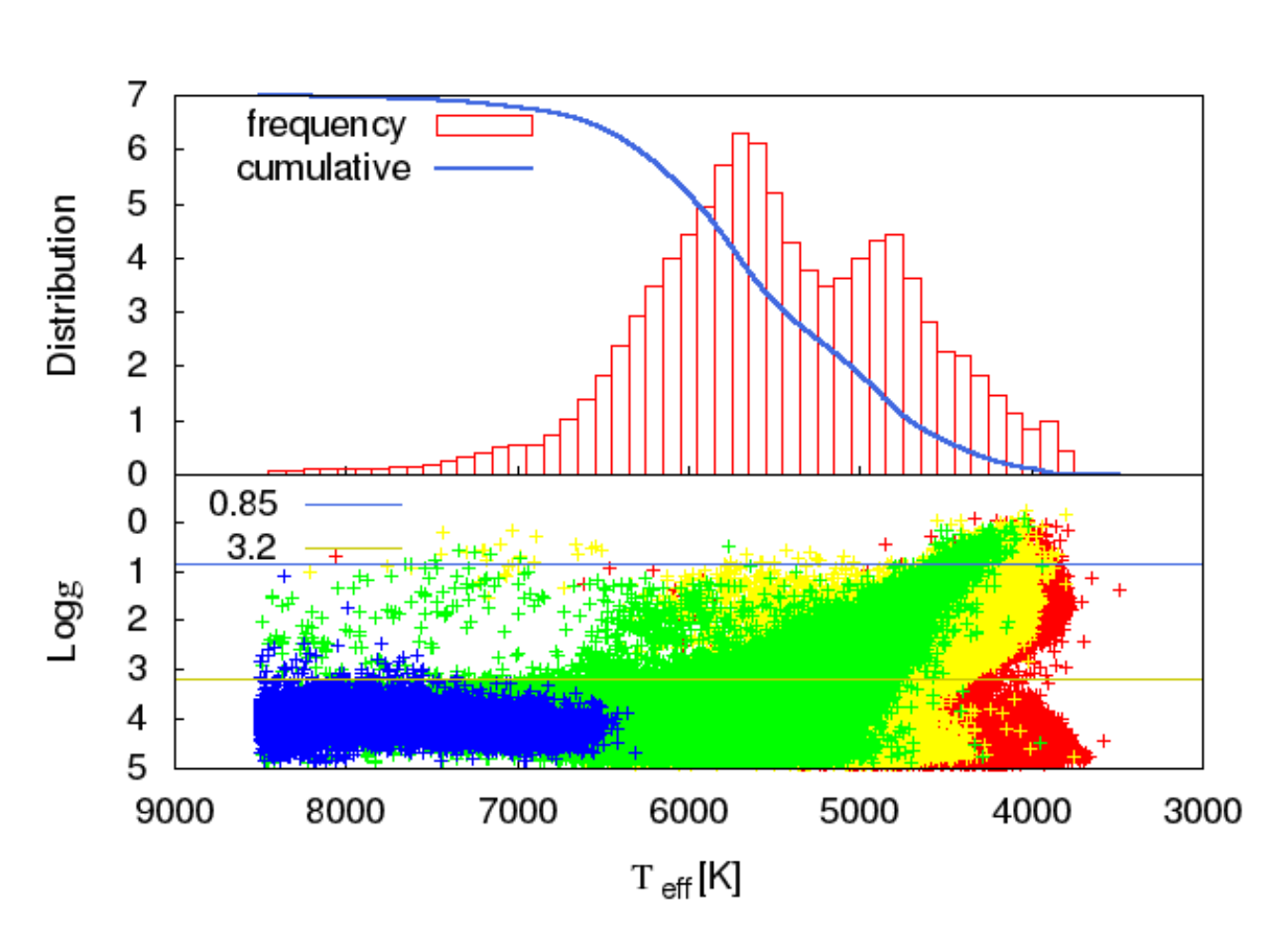}
\end{center}
\caption{The $T_{eff}$ to log$g$ figure for 6,478,063 AFGK type stars and the percentage histogram of the number of stars with $T_{eff}$ distribution. In the lower panel, the blue, green, yellow, and red pluses represent the A, F, G, and K type stars respectively. In the upper panel, the width of each histogram bar is set to 100\,K. The cumulative total number of stars is set to 7 at ordinate.}
\end{figure}

\begin{figure}
\begin{center}
\includegraphics[width=10.5cm,angle=0]{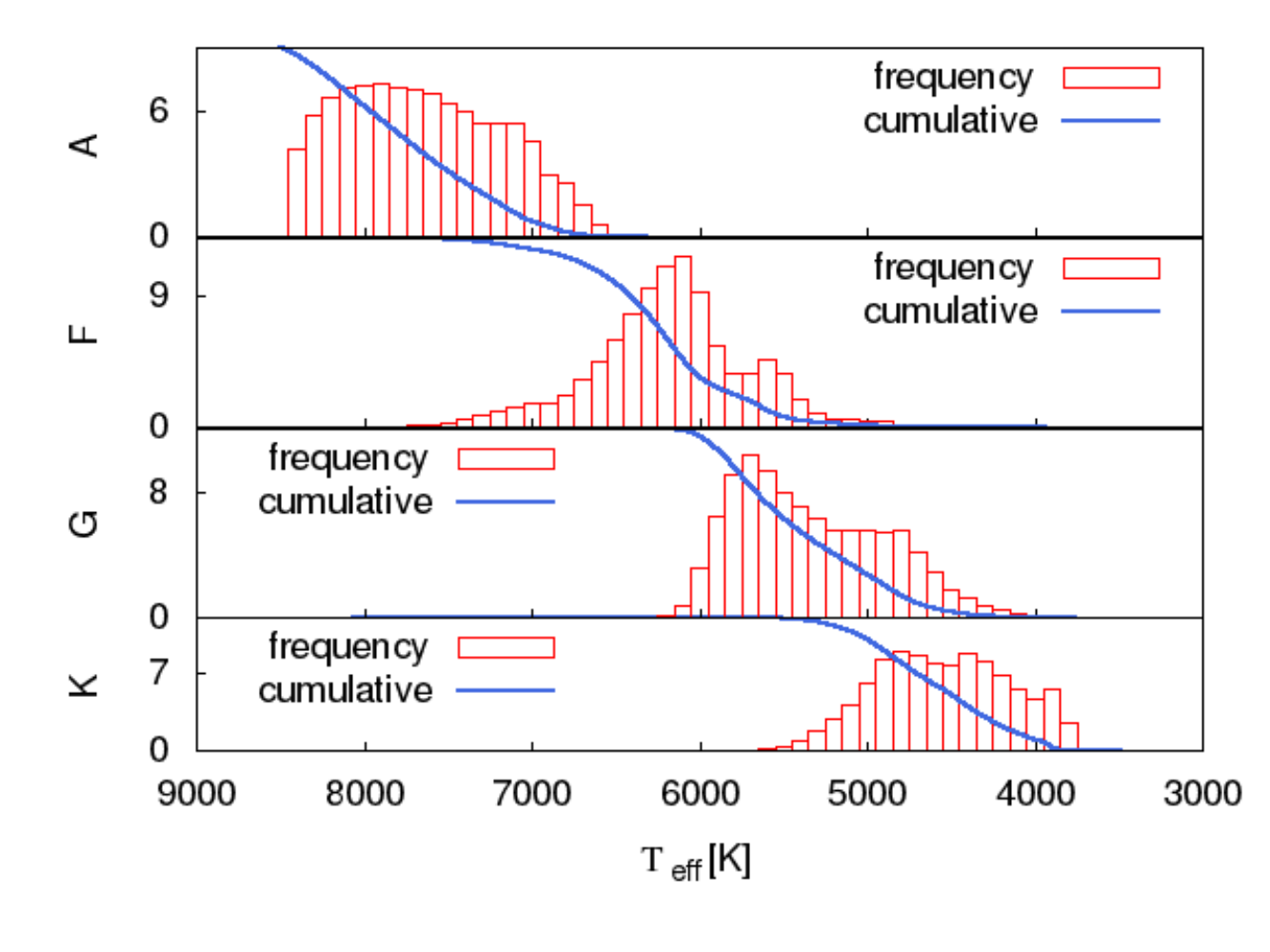}
\end{center}
\caption{The percentage histogram of the number of stars with $T_{eff}$ distribution for A, F, G, and K type stars respectively. The width of each histogram bar is set to 100\,K.}
\end{figure}

\begin{figure}
\begin{center}
\includegraphics[width=10.5cm,angle=0]{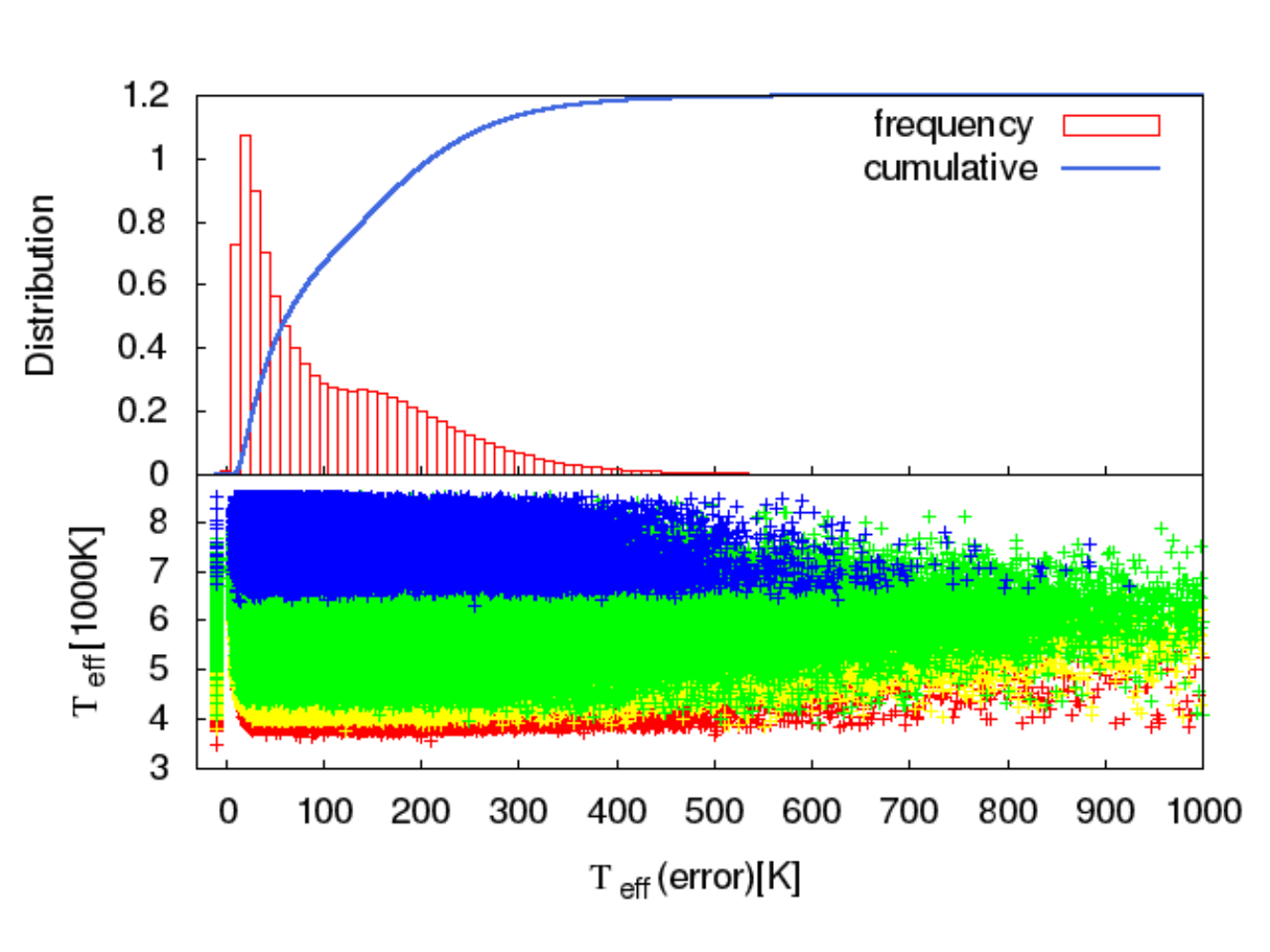}
\end{center}
\caption{The error of $T_{eff}$ to $T_{eff}$ figure for 6,478,063 AFGK type stars and the percentage histogram of the number of stars with the error of $T_{eff}$ distribution. The width of each histogram bar is set to 10\,K. The cumulative total number of stars is set to 1.2 at ordinate. There are 81\% of stars with the error of $T_{eff}$ less than 200\,K.}
\end{figure}

\begin{figure}
\begin{center}
\includegraphics[width=10.5cm,angle=0]{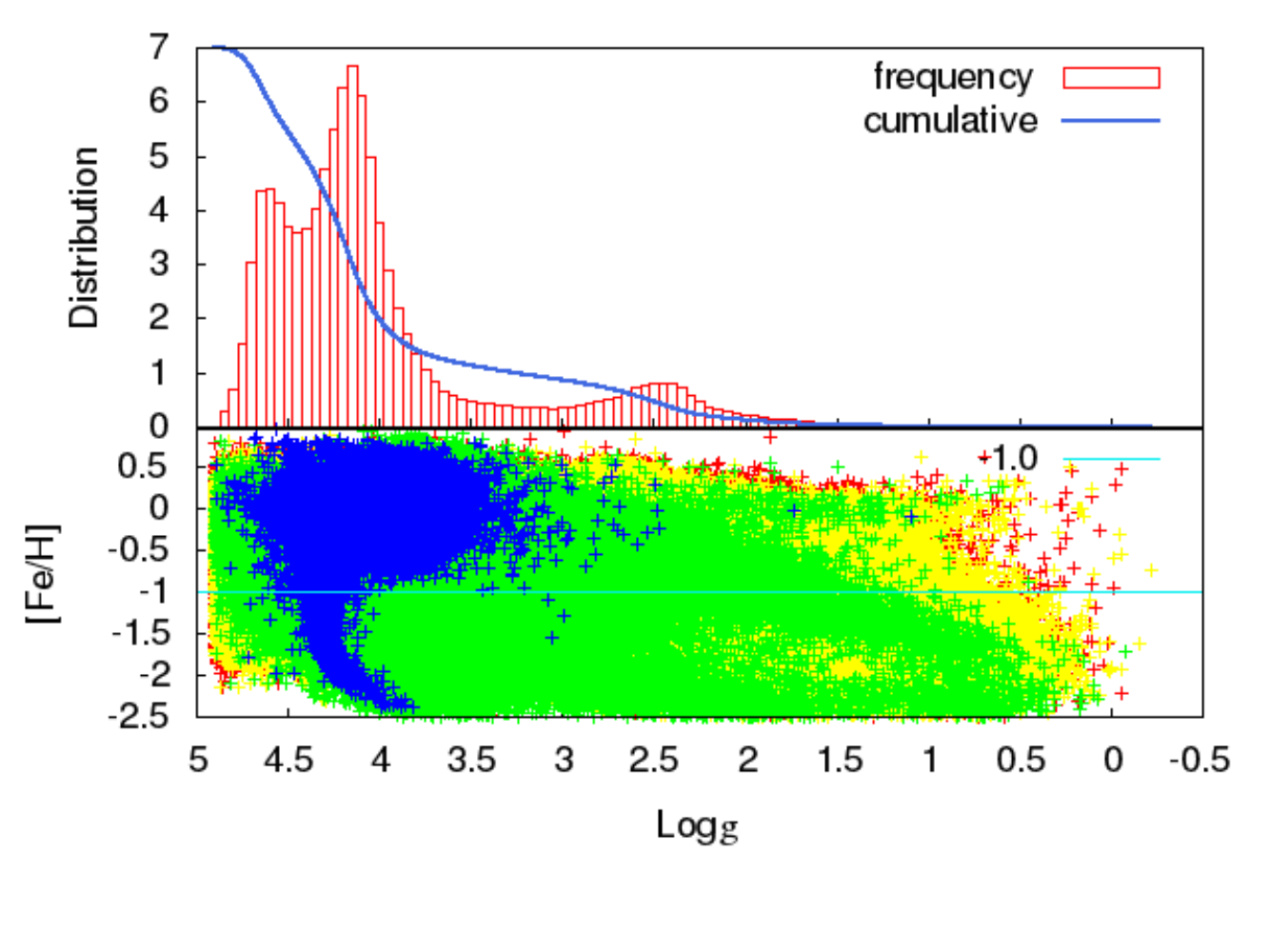}
\end{center}
\caption{The log$g$ to [Fe/H] figure for 6,478,063 AFGK type stars and the percentage histogram of the number of stars with log$g$ distribution. The width of each histogram bar is set to 0.05. The cumulative total number of stars is set to 7 at ordinate.}
\end{figure}

\begin{figure}
\begin{center}
\includegraphics[width=10.5cm,angle=0]{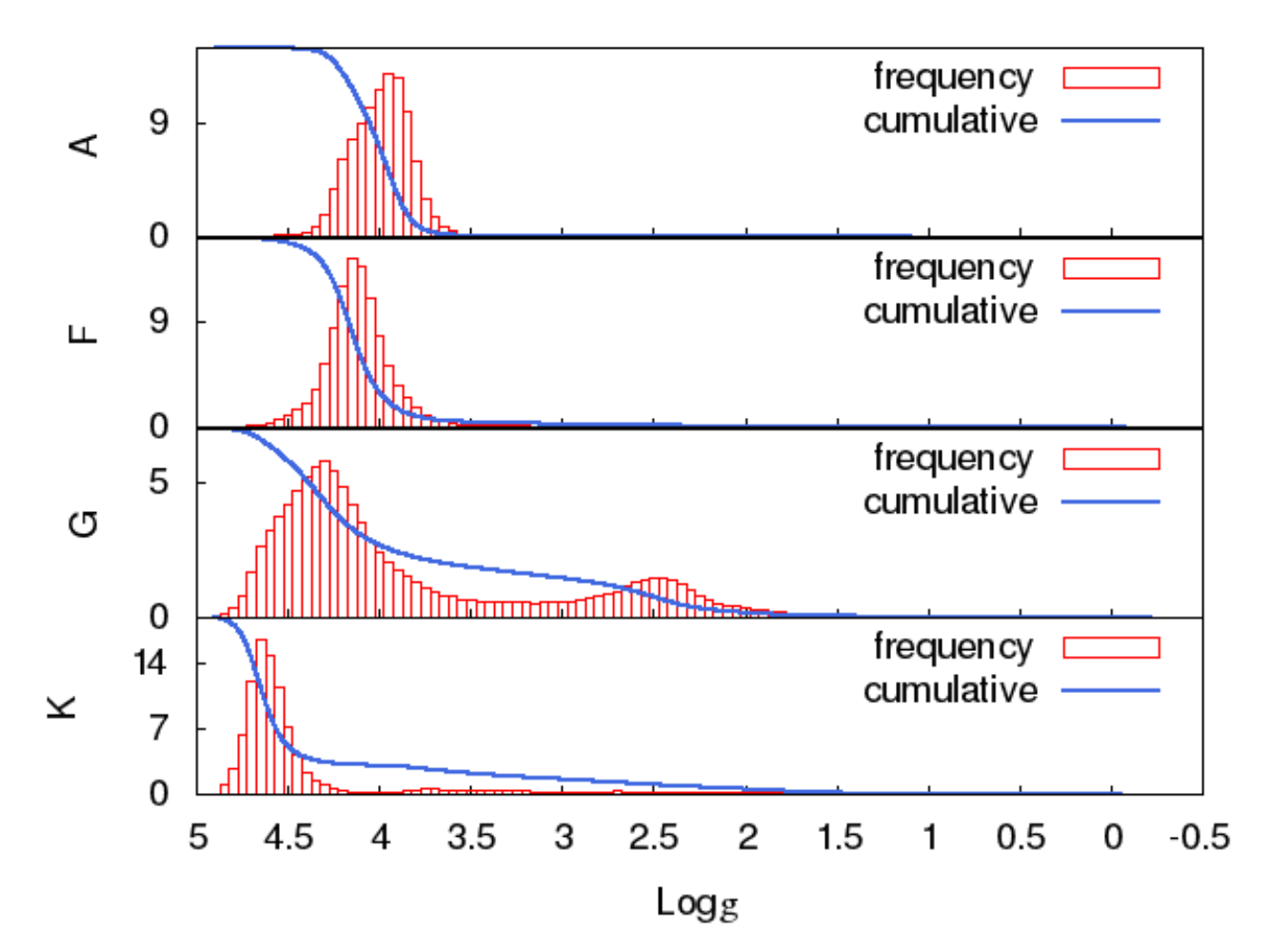}
\end{center}
\caption{The percentage histogram of the number of stars with log$g$ distribution for A, F, G, and K type stars respectively. The width of each histogram bar is set to 0.05.}
\end{figure}

\begin{figure}
\begin{center}
\includegraphics[width=10.5cm,angle=0]{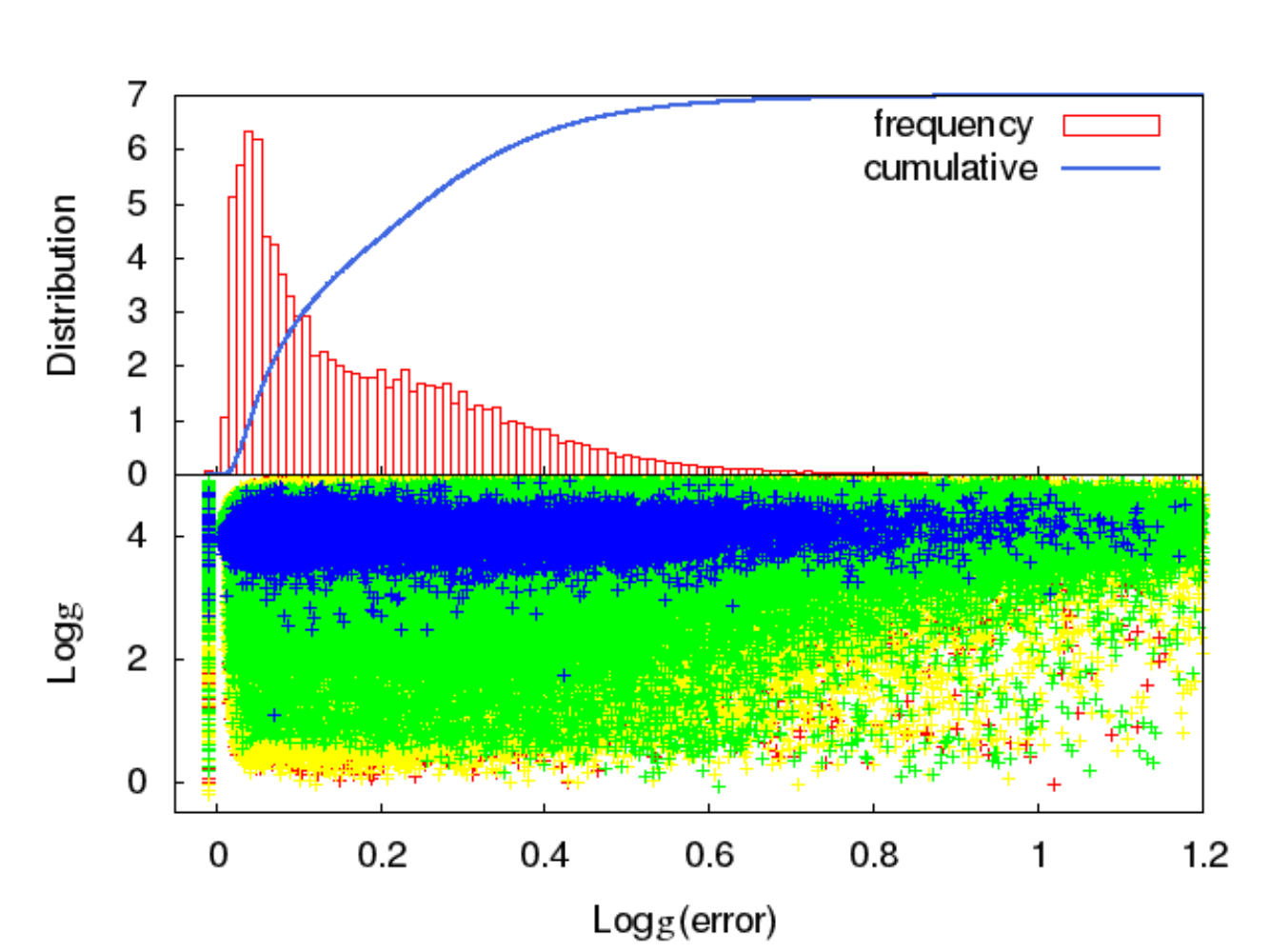}
\end{center}
\caption{The error of log$g$ to log$g$ figure for 6,478,063 AFGK type stars and the percentage histogram of the number of stars with the error of log$g$ distribution. The width of each histogram bar is set to 0.01. The cumulative total number of stars is set to 7.0 at ordinate. There are 79\% of stars with the error of log$g$ less than 0.3.}
\end{figure}

\begin{figure}
\begin{center}
\includegraphics[width=10.5cm,angle=0]{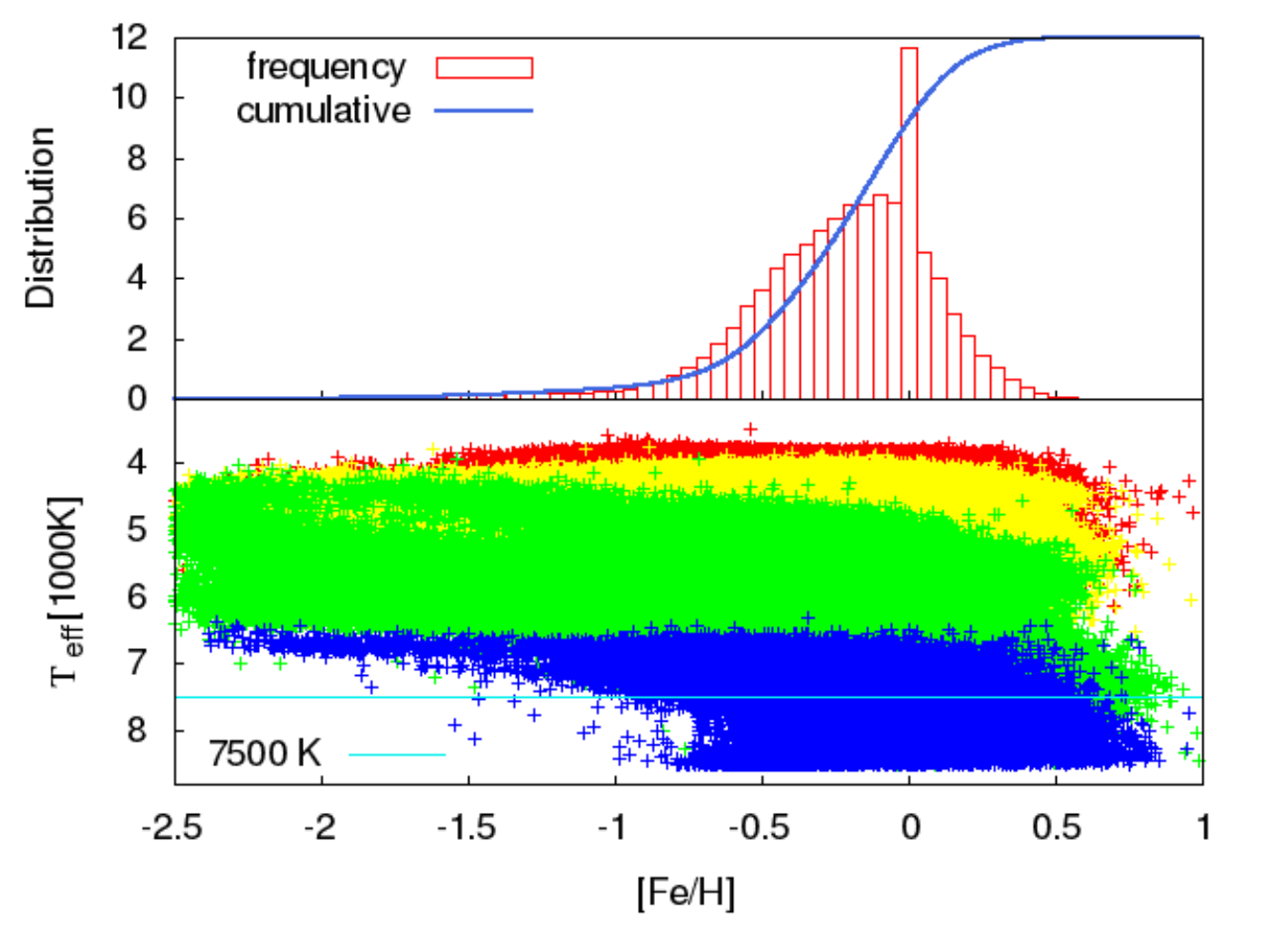}
\end{center}
\caption{The [Fe/H] to $T_{eff}$ figure for 6,478,063 AFGK type stars and the percentage histogram of the number of stars with [Fe/H] distribution. The width of each histogram bar is set to 0.05. The cumulative total number of stars is set to 12 at ordinate.}
\end{figure}

\begin{figure}
\begin{center}
\includegraphics[width=10.5cm,angle=0]{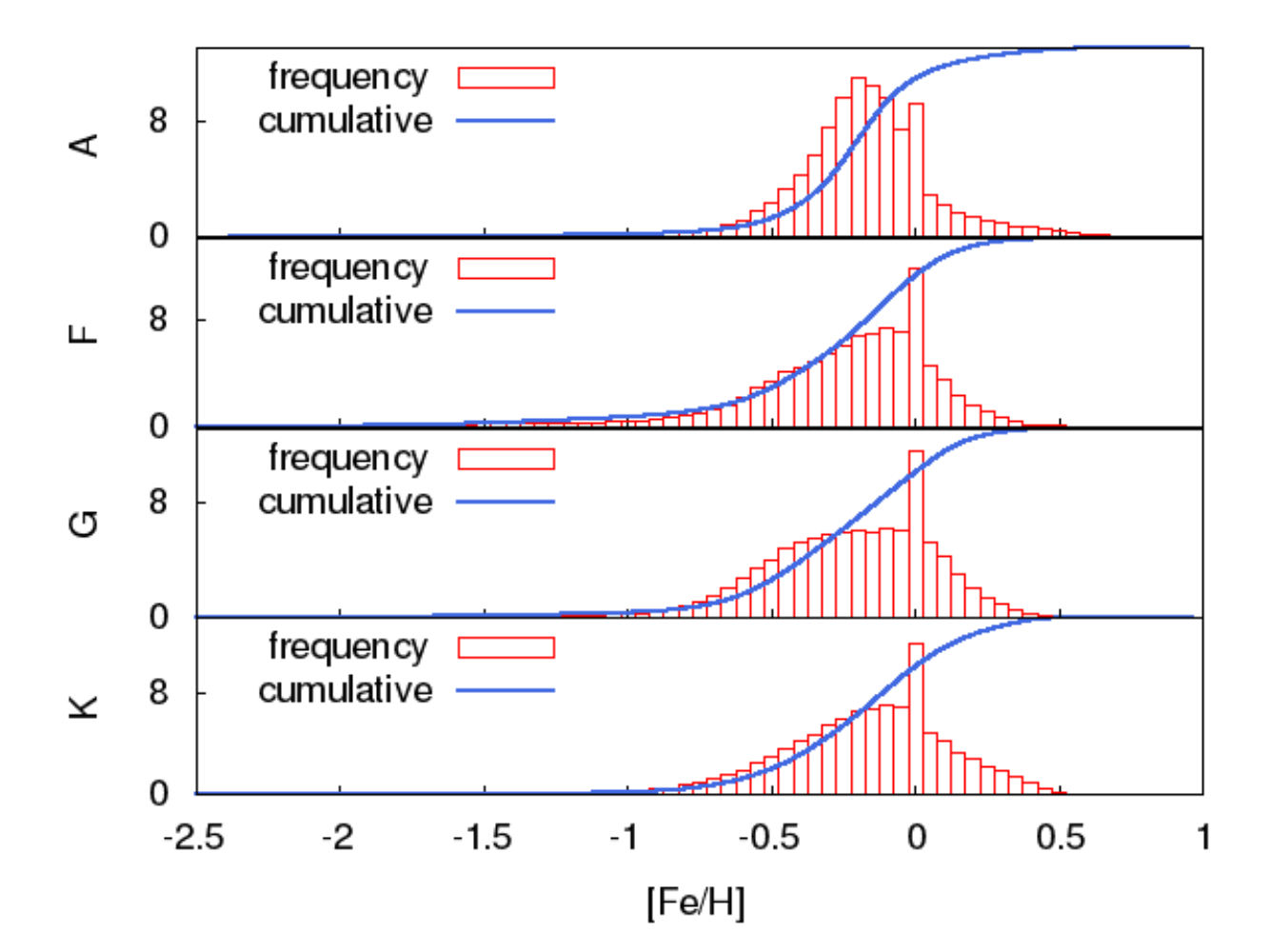}
\end{center}
\caption{The percentage histogram of the number of stars with [Fe/H] distribution for A, F, G, and K type stars respectively. The width of each histogram bar is set to 0.05.}
\end{figure}

\begin{figure}
\begin{center}
\includegraphics[width=10.5cm,angle=0]{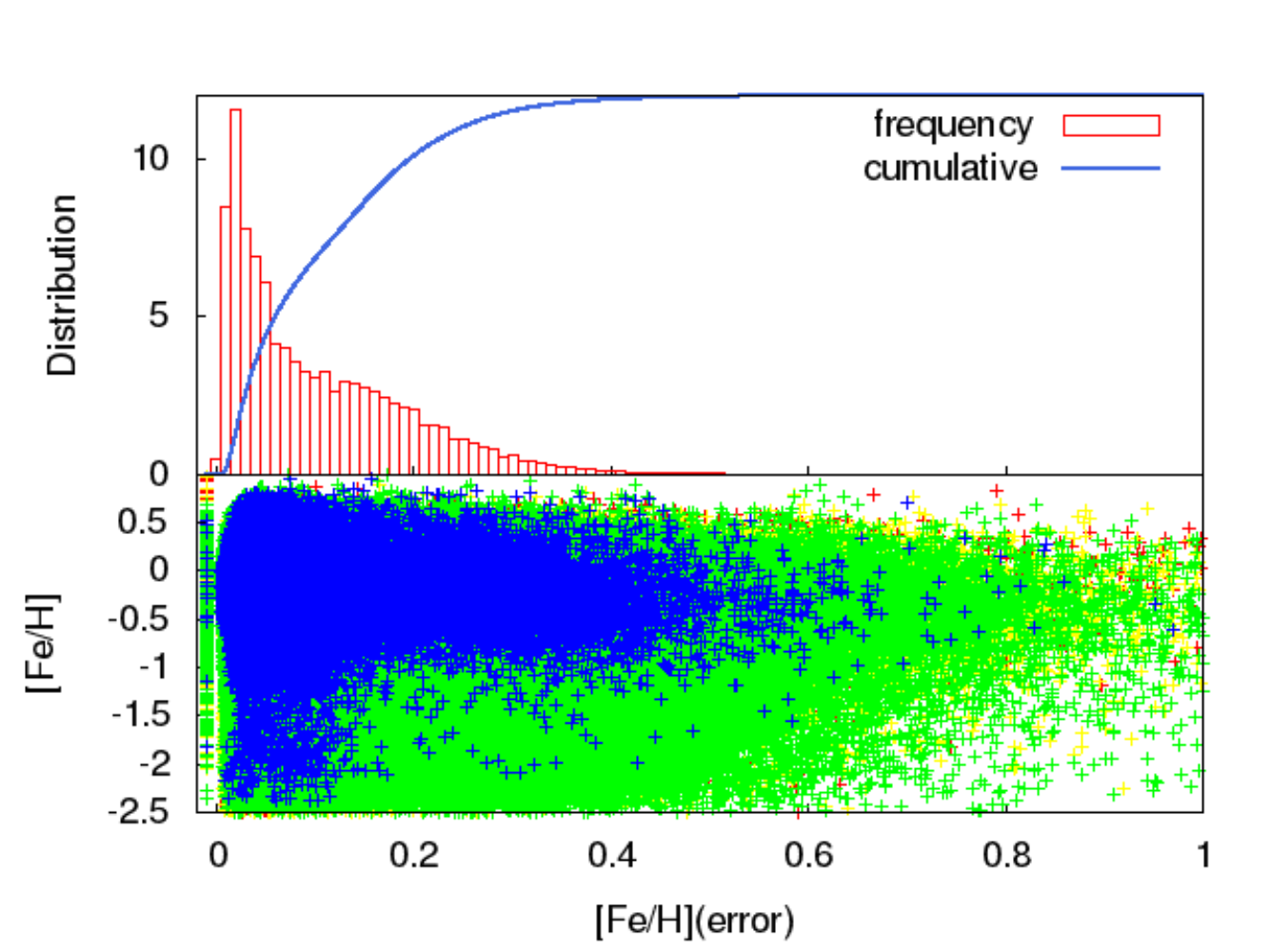}
\end{center}
\caption{The error of [Fe/H] to [Fe/H] figure for 6,478,063 AFGK type stars and the percentage histogram of the number of stars with the error of [Fe/H] distribution. The width of each histogram bar is set to 0.01. The cumulative total number of stars is set to 12 at ordinate. There are 84\% of stars with the error of [Fe/H] less than 0.2.}
\end{figure}

The 6,478,063 AFGK tpye stars contain 100,468 A type stars, 1,983,821 F type stars, 3,249,746 G type stars, and 1,144,028 K type stars. These spectra are published by LAMOST DR8 v1.0 with signal to noise ratio in g band larger than 15 in bright nights and 6 in dark nights. The wavelength range of the spectra is from 3,690 {\AA} to 9,100 {\AA}. The resolution of the spectra is 1,800 at the 5,500 {\AA} (Stoughton et al. 2002, Abazajian et al. 2003). Based on the LAMOST Stellar Parameter pipeline (LASP) (Wu et al. 2014, Luo et al. 2015), the effective temperatures ($T_{eff}$), surface gravities (log$g$), and metallicities ([Fe/H]) are calculated and published. With these statistical parameters, we perform a preliminary research work.

In Fig. 1, we show the $T_{eff}$ to log$g$ figure for 6,478,063 AFGK type stars and the percentage histogram of the number of stars with $T_{eff}$ distribution. In the lower panel, the blue, green, yellow, and red pluses represent the A, F, G, and K type stars respectively. The log$g$ values of most A type stars are greater than 3.2. The values of log$g$ have a gap area around 3.2 at the red end of $T_{eff}$ around 4,000\,K. Most stars with log$g$ less than 0.85 have red $T_{eff}$ values, around 4,000\,K. In Fig. 2, we show the percentage histogram of the number of stars with $T_{eff}$ distribution for A, F, G, and K type stars respectively. In Fig 2, we can see that most of the A, F, G, and K type stars are consistent with the Harvard system of spectral classification of stars. The spectral parameters are self consistent. For the parameter errors, the error of $T_{eff}$ for 1,268 stars can not be calculated and is marked as -9999 in the original data. We set these -9999 to -10 and draw Fig. 3. Figure 3 is the error of $T_{eff}$ to $T_{eff}$ figure for 6,478,063 AFGK type stars and the percentage histogram of the number of stars with the error of $T_{eff}$ distribution. There are 81\% of stars with the error of $T_{eff}$ less than 200\,K.

In Fig. 4, we show the log$g$ to [Fe/H] figure for 6,478,063 AFGK type stars and the percentage histogram of the number of stars with log$g$ distribution. The ordinate of the lower panel is the [Fe/H], the logarithm value of a stellar average heavy-element abundance relative to the Sun, not only iron content. By checking the upper panel and the original data, we obtain that there are 5,276 (445 F type, 4,109 G type, and 722 K type) stars with log$g$ less than and equal to 0.85. In the lower panel, most of the stars with log$g$ less than 0.85 have [Fe/H] in a range from -2.5 to -1.0, corresponding metal abundance Z from 0.000063 to 0.002. The metal abundance for the Sun is Z = 0.02. Most of the stars are metal poor stars, approximately belonging to stellar population II, when log$g$ is less than 0.85. In addition, in the upper panel, there is one peak around log$g$ = 4.2 and the other peak around log$g$ = 2.4. There is a valley around log$g$ = 3.2. In Fig. 5, we show the percentage histogram of the number of stars with log$g$ distribution for A, F, G, and K type stars respectively. The log$g$ values of K type stars are slightly larger. The peak around log$g$ = 2.4 in Fig. 4 is mainly caused by G type stars, as shown in Fig. 5. In Fig. 6, we show the error of log$g$ to log$g$ for 6,478,063 AFGK type stars and the percentage histogram of the number of stars with the error of log$g$ distribution. The error of log$g$ for 4,139 stars can not be calculated and is marked
as -0.01 in Fig. 6. There are 79\% of stars with the error of log$g$ less than 0.3.

In Fig. 7, we show the [Fe/H] to $T_{eff}$ figure for 6,478,063 AFGK type stars and the percentage histogram of the number of stars with [Fe/H] distribution. The stars with the same metal abundance as the Sun, [Fe/H] = 0.0, are the most and account for 6\% in the upper panel of Fig. 7. Most of the stars with $T_{eff}$ higher than 7,500\,K have [Fe/H] in a range from -0.75 to 0.75, corresponding to the metal abundance Z from 0.0036 to 0.11. Namely, most of the stars are metal rich stars, approximately belonging to stellar population I, when $T_{eff}$ is higher than 7,500\,K. In fact, in the upper panel, we can see that the number of stars with extremely poor metal abundances is relatively small. By checking the upper panel of Fig. 1 and the original data, we obtain that there are 78,141 stars with $T_{eff}$ greater than and equal to 7,500\,K. These 78,141 stars contain 65,842 A type stars, 12,295 F type stars, 3 G type stars, and 1 K type star. Most of these 78,141 stars are hot A type stars. The log$g$ values of them are greater than 3.5 according to the top panel of Fig. 5. In Fig. 8, we show the percentage histogram of the number of stars with [Fe/H] distribution for A, F, G, and K type stars respectively. Except for the A type stars, the metal abundances of F, G, and K type stars all have a peak value at [Fe/H] = 0.0. However, the metal abundances among A, F, G, and K type stars have a wide distribution. The metal abundance is an important parameter in stellar physics. In a fine research work, we can not simply replace the metal abundance of other stars with the metal abundance of the Sun. In Fig. 9, we show the error of [Fe/H] to [Fe/H] for 6,478,063 AFGK type stars and the percentage histogram of the number of stars with the error of [Fe/H] distribution. The error of [Fe/H] for 1,095 stars can not be calculated and is marked as -0.01 in Fig. 9. There are 84\% of stars with the error of [Fe/H] less than 0.2.

\section{The preliminary analysis of observation results based on evolutionary models}

\begin{figure}
\begin{center}
\includegraphics[width=10.5cm,angle=0]{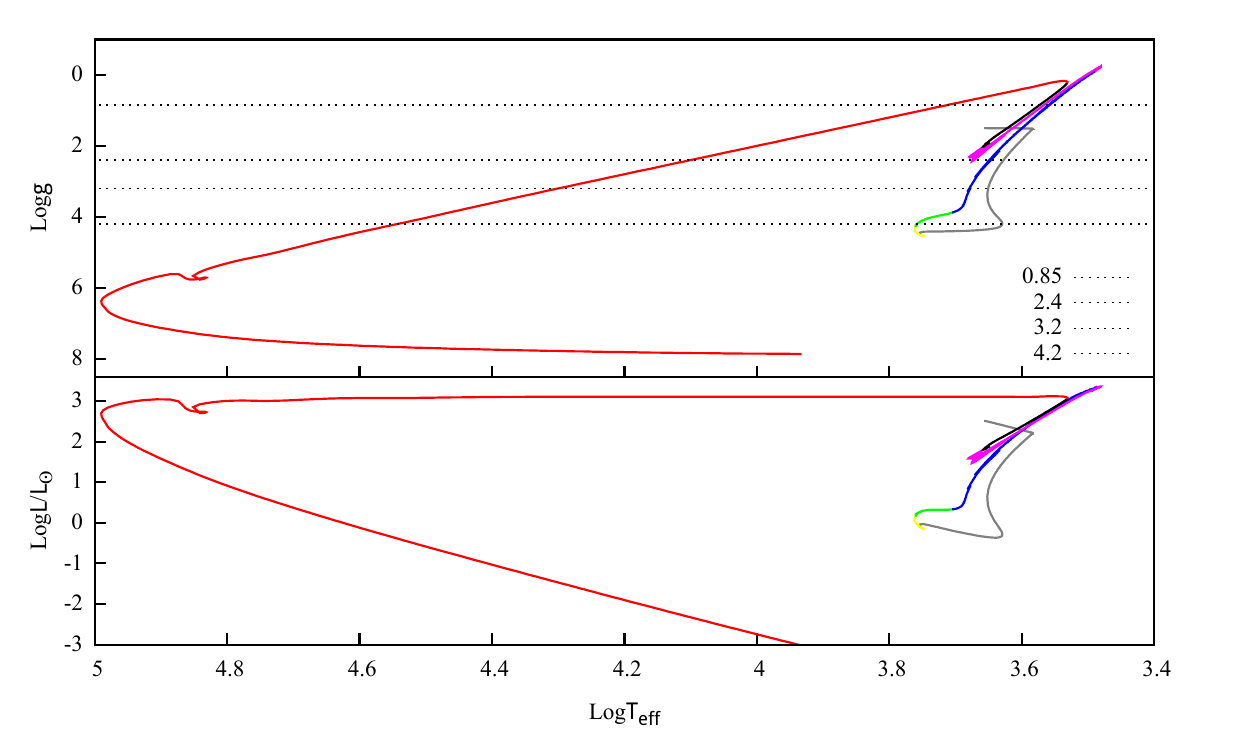}
\end{center}
\caption{The H-R and corresponding log$T_{eff}$ to log$g$ diagram for 1.0\,$M_{\bigodot}$ stellar evolution. The gray, yellow, blue, pink, black, and red line represents the pre-MS, MS, RGB, HB, AGB, and PN-WD respectively. The green line represents the transition from the MS stage to the RGB stage.}
\end{figure}

\begin{figure}
\begin{center}
\includegraphics[width=10.5cm,angle=0]{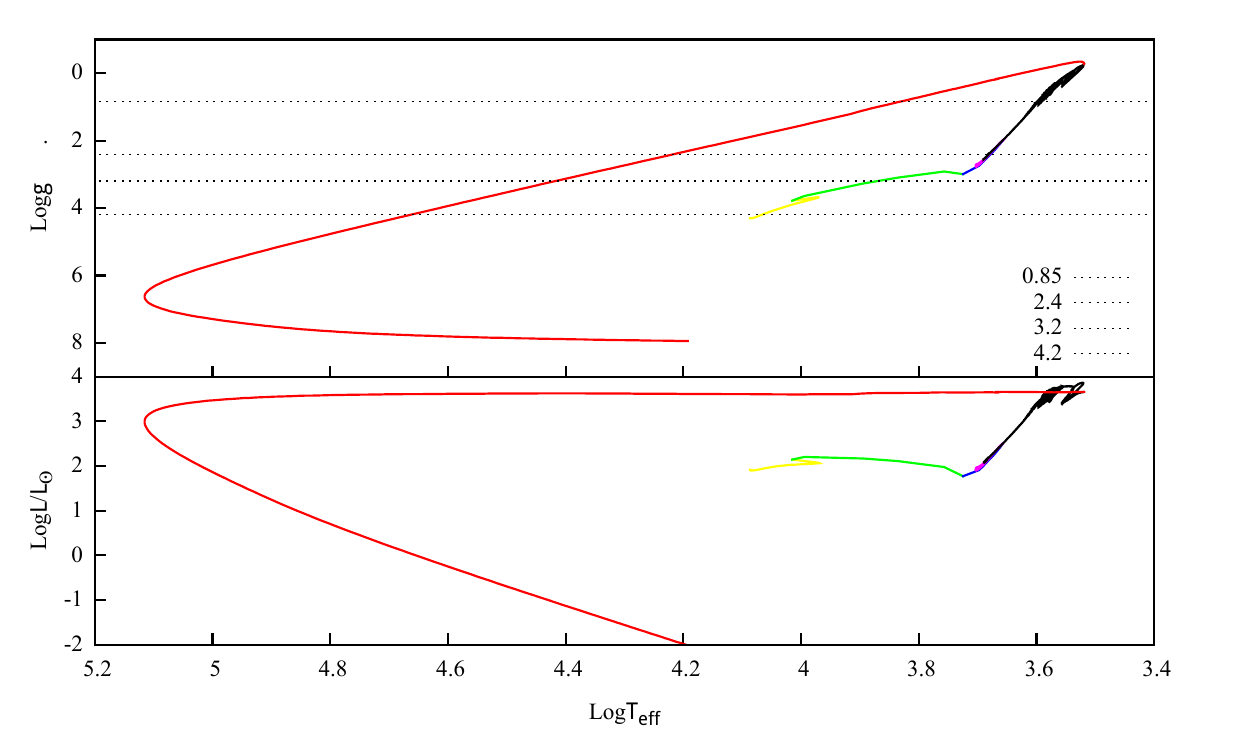}
\end{center}
\caption{The H-R and corresponding log$T_{eff}$ to log$g$ diagram for 3.0\,$M_{\bigodot}$ stellar evolution. The yellow, blue, pink, black, and red line represents the MS, RGB, HB, AGB, and PN-WD respectively. The green line represents the gap area in the H-R diagram.}
\end{figure}

\begin{figure}
\begin{center}
\includegraphics[width=10.5cm,angle=0]{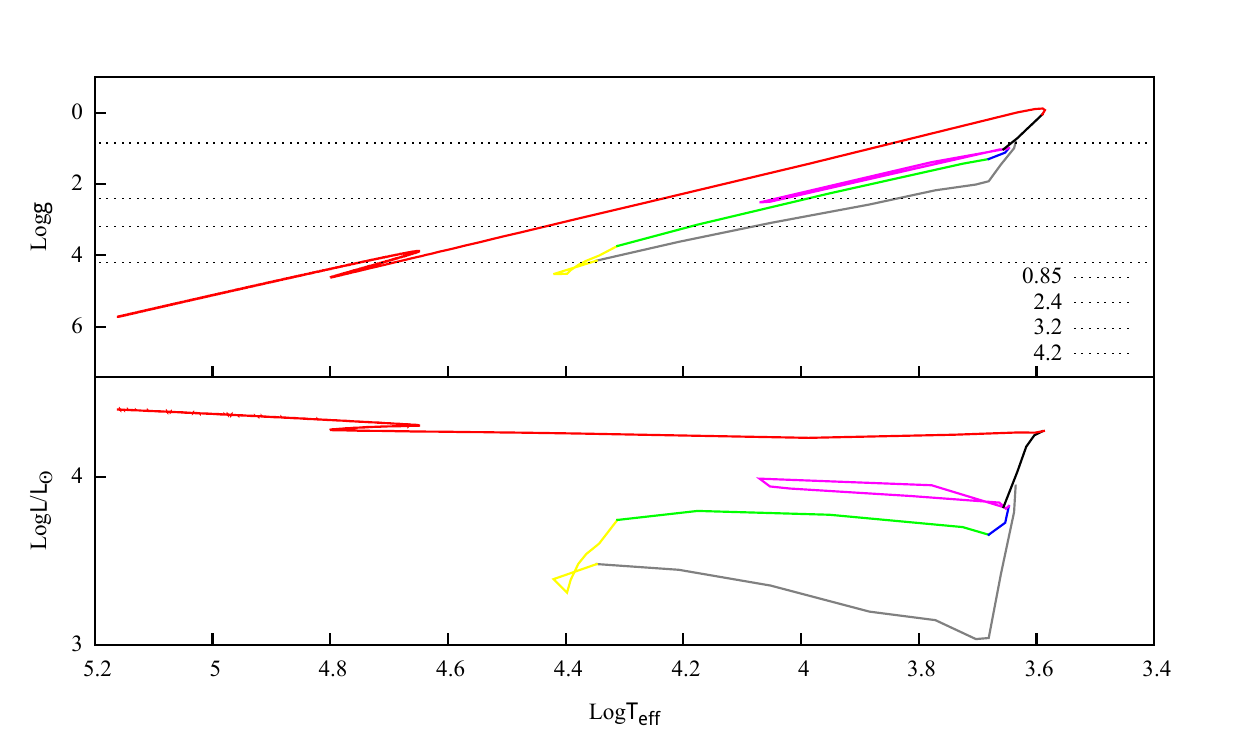}
\end{center}
\caption{The H-R and corresponding log$T_{eff}$ to log$g$ diagram for 7.0\,$M_{\bigodot}$ stellar evolution. The gray, yellow, blue, pink, black, and red line represents the pre-MS, MS, RGB, HB, AGB, and PN respectively. The green line represents the gap area in the H-R diagram.}
\end{figure}

In this section, we do some stellar evolutions and perform a preliminary analysis of the observation results based on the evolutionary models. In Fig. 10, we show a Hertzsprung-Russell (H-R) diagram and the corresponding log$T_{eff}$ to log$g$ diagram for 1.0\,$M_{\bigodot}$ stellar evolution based on stellar evolution code \texttt{MESA} (Paxton et al. 2011), the module of 1M\_pre\_ms\_to\_wd. \texttt{MESA} is a powerful stellar evolution code and it is the abbreviation of the modules for experiments in stellar astrophysics (Paxton et al. 2011). The equation of state tables are from the OPAL EOS tables (Rogers \& Nayfonov 2002), SCVH tables (Saumon et al. 1995) for lower temperatures and densities, HELM tables (Timmers \& Swesty 2000) and PC tables (Potekhin \& Chabrier 2010) for complete ionization regions. The opacity tables are from Iben (1975) for non-degenerate electrons, Yakovlev \& Urpin (1980) for degenerate electrons, Ferguson et al. (2005) and Iglesias \& Rogers (1993, 1996) for radiative opacities, and Cassisi et al. (2007) for the combinations between radiative opacities and electron conduction opacities. The thermonuclear burning rates are from Angulo et al. (1999) and Caughlan \& Fowler (1988). For more details about \texttt{MESA}, see Paxton et al. (2011) and their follow-up papers.

In Fig. 10, the gray line represents the pre main sequence (pre-MS) stage. The ignition of nuclear reaction (log$T_{c}$ [K] = 7.14 for the model) marks that the star has entered the zero age main sequence stage. The yellow line represents the main sequence (MS) stage. The green line represents the transition from the MS stage to the red giant branch (RGB) stage. At this stage, the central hydrogen burning ends and a hydrogen shell burning appears at the edge of the central helium core. The blue line represents the RGB stage. The pink line represents the core helium burning stage. These stars at the core helium burning stage are called horizontal branch (HB) stars (Faulkner 1966). The HB stars are observed in the color-magnitude diagram of globular clusters, such as M15 (Durrell et al. 1993). When the central helium core burning ends, a hydrogen burning shell coexists with a helium burning shell below it. The star enters the asymptotic giant branch (AGB) stage, as shown by the black line in Fig. 10. The AGB stars have a hydrogen burning shell and a helium burning shell, while the RGB stars only have a hydrogen burning shell. The red line represents the planetary nebula (PN) stage and white dwarf (WD) stage. The lower panel of Fig. 10 is the H-R diagram. The upper panel of Fig. 10 is the corresponding log$T_{eff}$ to log$g$ diagram. The dotted lines in the upper panel of Fig. 10 are to facilitate the identification of the positions where log$g$ are taken as 0.85, 2.4, 3.2, and 4.2. These values of log$g$ are typical values in Sect. 2.

In Fig. 11,  we show a H-R diagram and the corresponding log$T_{eff}$ to log$g$ diagram for 3.0\,$M_{\bigodot}$ stellar evolution based on \texttt{MESA}, the module of mk\_co\_wd. In Fig. 12,  we show a H-R diagram and the corresponding log$T_{eff}$ to log$g$ diagram for 7.0\,$M_{\bigodot}$ stellar evolution based on \texttt{MESA}, the module of 7M\_prems\_to\_AGB. The input physics are default for the three stellar evolutions. Stars with masses smaller than $\sim$ 2.2\,$M_{\bigodot}$ are called small mass stars. The central helium core is electron degenerate before ignition. Stars with masses from $\sim$ 2.2\,$M_{\bigodot}$ to $\sim$ 9.0\,$M_{\bigodot}$ are called medium mass stars. The central carbon/oxygen core is electron degenerate. Stars with masses larger than $\sim$ 9.0\,$M_{\bigodot}$ are called large mass stars. The central core is nondegenerate before it becomes an iron core. The large mass stars are usually O type stars and B type stars, which are not evolved in this paper. We evolve a small mass star and two medium mass stars as examples to do a preliminary analysis and research work. At the stage of the green lines in Fig. 11 and 12, the helium core inside the hydrogen burning shell shrinks rapidly, the envelope outside the hydrogen burning shell expands rapidly. The time scales are thermodynamic and the stage is called the gap area in the H-R diagram. The stars at this stage are too short time scales to be observed. More details about stellar structure and evolution results can be referred to the tables of Schalle et al. (1992).

In Fig. 10, 11 and 12, the stars with log$g$ less than 0.85 and $T_{eff}$ around 4,000\,K (log4,000 = 3.60) are RGB or AGB stars. Combined with Fig. 1 and Fig. 4, we can conclude that most RGB or AGB stars with log$g$ less than 0.85 have metal abundances Z being from 0.000063 to 0.002. They are metal poor stars, approximately belonging to stellar population II. The conclusion is very important for RGB and AGB stars. For example, when studying the asteroseismology of RGB stars with log$g$ less than 0.85, more consideration should be given to the evolved RGB star models with poor metal abundances. In the upper panel of Fig. 4, there are only 5,276 stars with log$g$ less than and equal to 0.85. The hydrogen burning stage accounts for nearly 90\% of the stellar lifetime and the helium burning stage accounts for nearly 10\% of the stellar lifetime. The central core burning process is the main process of stars. It is reasonable that among the stars, the stars with shell burning and log$g$ less than 0.85 account for a small proportion.

At the red end of the lower panel of Fig. 1, the value of log$g$ has a gap area around 3.2. In Fig. 4 (the upper panel) and Fig. 5 (the G type stars), there is a valley around log$g$ = 3.2. It corresponds to the RGB stars in Fig. 10 for stellar evolution of a small mass star and the gap area in the H-R diagram in Fig. 11 and 12 for stellar evolution of medium mass stars. The stars in the valley of Fig. 5 for G type stars are mainly from RGB stars for stellar evolutions of small mass stars. The theory of the gap area in H-R diagram for stellar evolutions of medium mass stars is consistent with the big data observation and statistical results. In the upper panel of Fig. 4, there is one peak around log$g$ = 4.2 and the other peak around log$g$ = 2.4. In Fig. 10, 11, and 12, the stars of log$g$ $\sim$ 4.2 corresponds to the MS stars. The stars of log$g$ $\sim$ 2.4 corresponds to the HB or RGB stars for stellar evolutions of small mass stars, the HB or RGB or AGB stars for stellar evolution of medium mass stars.

Combined with Fig. 1 and 7, most stars with $T_{eff}$ higher than 7,500\,K (log7,500 = 3.88) have log$g$ from 3.2 to 4.8 and metal abundance Z from 0.0036 to 0.11. Except for white dwarfs, most stars with such a large gravitational acceleration ($T_{eff}$ higher than 7,500\,K) in Fig. 10, 11 and 12 are MS stars. Therefore, most A type MS stars with $T_{eff}$ higher than 7,500\,K are metal rich stars, approximately belonging to stellar population I. It indicates that the metal abundances of the hot A type MS stars are not completely consistent with that of the Sun. The metal abundances, especially metal rich abundances, should be considered in the study of the hot A type MS stars.

\section{Discussion and conclusions}

\begin{figure}
\begin{center}
\includegraphics[width=10.5cm,angle=0]{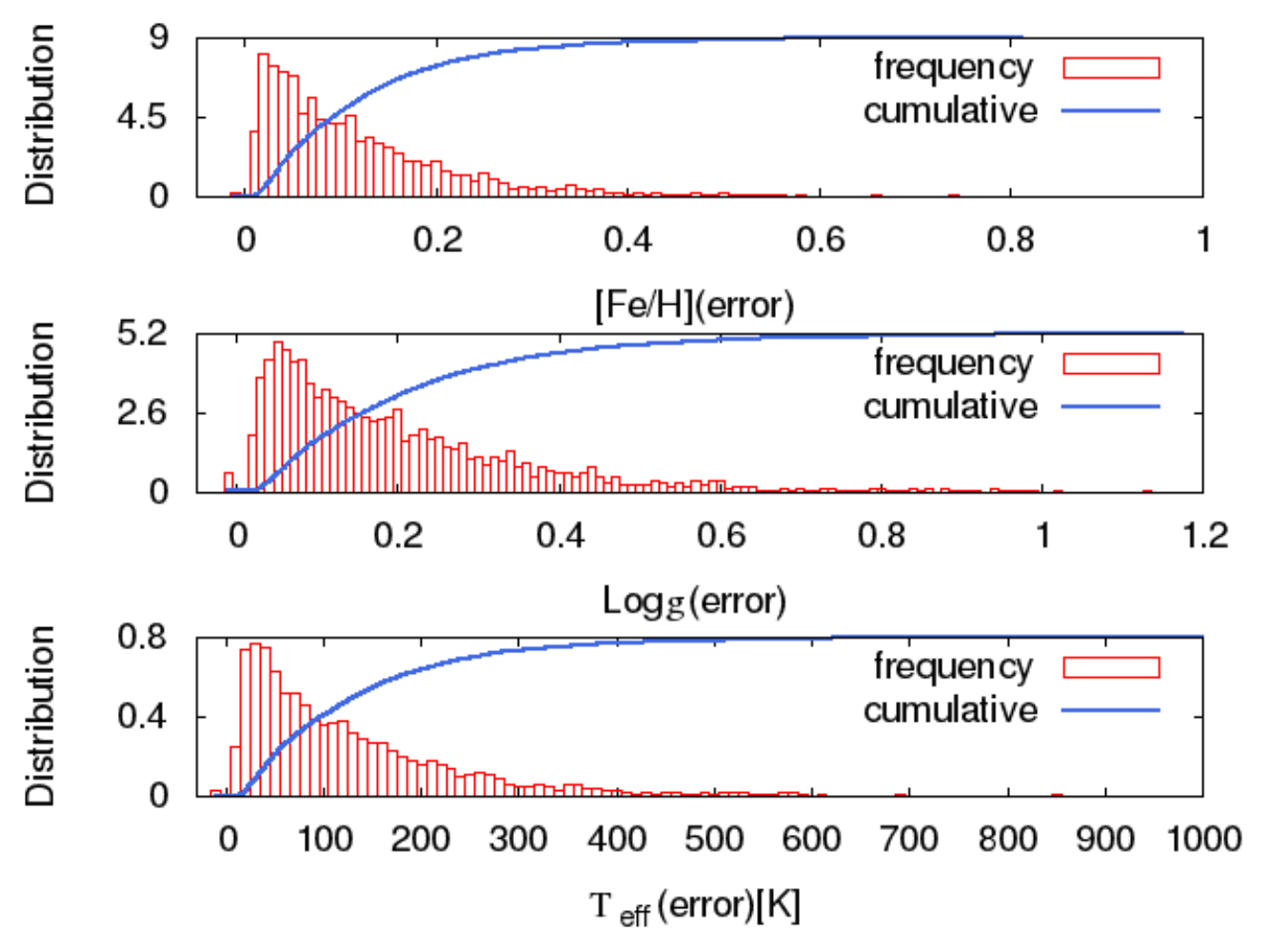}
\end{center}
\caption{The percentage histogram of the number of 5,276 stars (log$g$ $\leq$ 0.85) with the error of $T_{eff}$, log$g$, and [Fe/H] distribution. In the lower panel, there are 51\%, 79\%, and 92\% of stars with the error of $T_{eff}$ less than 100\,K, 200\,K, and 300\,K respectively, including 15 negative (no error of $T_{eff}$) values. In the middle panel, there are 33\%, 61\%, and 78\% of stars with the error of log$g$ less than 0.1, 0.2, and 0.3 respectively, including 34 negative (no error of log$g$) values. In the upper panel, there are 53\%, 82\%, and 93\% of stars with the error of [Fe/H] less than 0.1, 0.2, and 0.3 respectively, including 14 negative (no error of [Fe/H]) values.}
\end{figure}

\begin{figure}
\begin{center}
\includegraphics[width=10.5cm,angle=0]{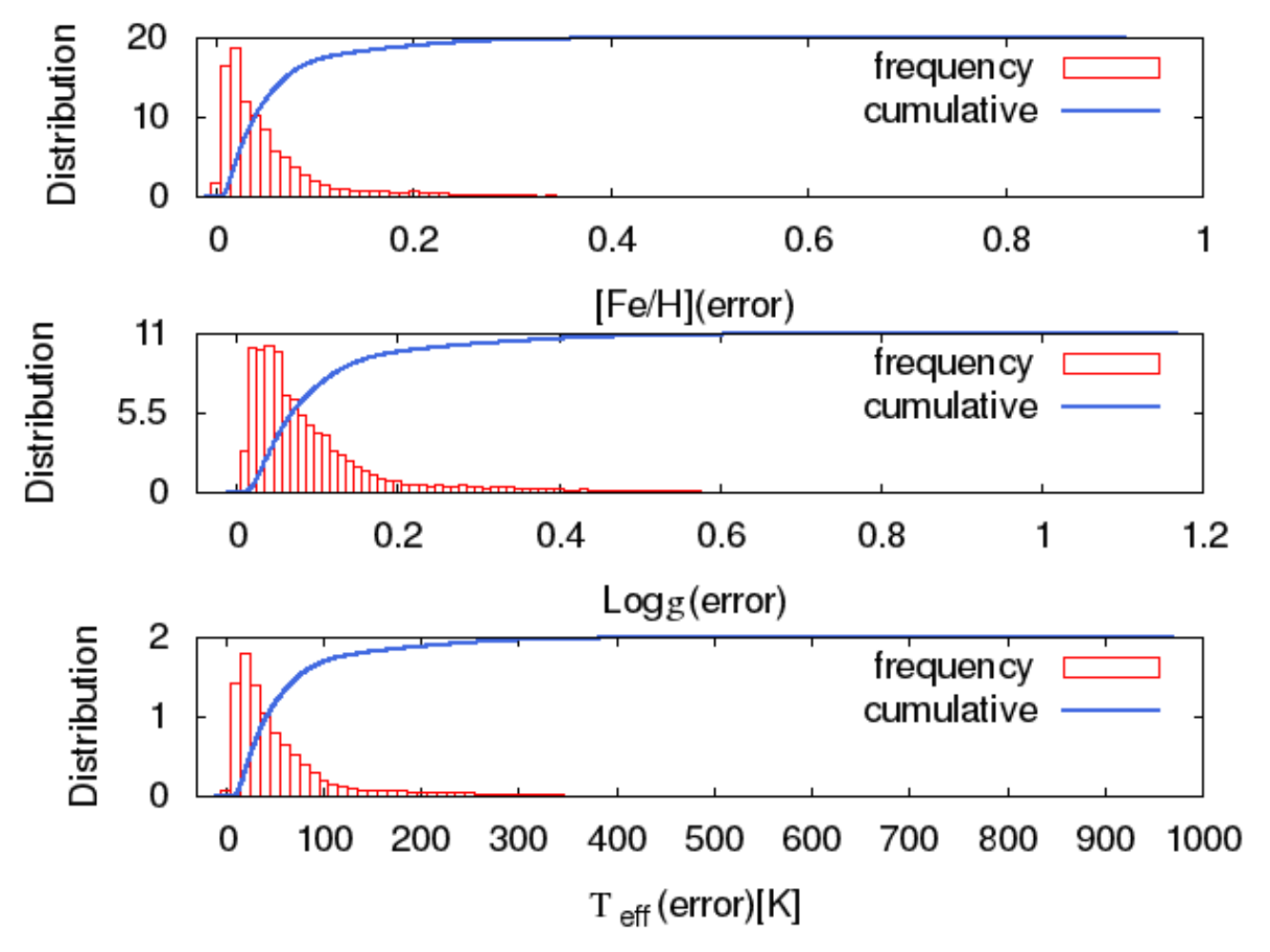}
\end{center}
\caption{The percentage histogram of the number of 78,141 stars ($T_{eff}$ $\geq$ 7,500\,K) with the error of $T_{eff}$, log$g$, and [Fe/H] distribution. In the lower panel, there are 84\%, 94\%, and 98\% of stars with the error of $T_{eff}$ less than 100\,K, 200\,K, and 300\,K respectively. In the middle panel, there are 66\%, 88\%, and 93\% of stars with the error of log$g$ less than 0.1, 0.2, and 0.3 respectively. In the upper panel, there are 85\%, 95\%, and 99\% of stars with the error of [Fe/H] less than 0.1, 0.2, and 0.3 respectively.}
\end{figure}

In this paper, we download the data of LAMOST DR8 low resolution catalog AFGK type stars and do a basic analysis based on the $T_{eff}$, log$g$, and [Fe/H] values of 6,478,063 stars. We evolve 1.0\,$M_{\bigodot}$, 3.0\,$M_{\bigodot}$ and 7.0\,$M_{\bigodot}$ stars from the pre-MS stage or MS stage to the PN stage or WD stage by \texttt{MESA}. Combined with the observed statistical data and model calculation data, we draw some basic conclusions preliminarily. In the lower panels of Fig. 3, 6, and 9, we can see that the error ranges of $T_{eff}$, log$g$, and [Fe/H] are large. However, in the corresponding upper panels, there are 81\%, 79\%, and 84\% of stars with the errors of $T_{eff}$, log$g$, and [Fe/H] less than 200\,K, 0.3, and 0.2 respectively. Therefore, within a certain error range, we can draw some conclusions.

Most RGB and AGB stars with log$g$ less than 0.85 have low metal abundance. This provides a guidance for the study of metal abundance input of RGB stars and AGB stars. In the future work, we will study the effects of different metal abundances on the asteroseismology of the RGB stars. There are 5,276 stars with log$g$ $\leq$ 0.85. we show the percentage histogram of the number of the 5,276 stars with the error of $T_{eff}$, log$g$, and [Fe/H] distribution in Fig. 13. There are total 32\%, 60\%, and 78\% of stars with the error of $T_{eff}$ $\leq$ 100\,K / log$g$ $\leq$ 0.1 / [Fe/H] $\leq$ 0.1, the error of $T_{eff}$ $\leq$ 200\,K / log$g$ $\leq$ 0.2 / [Fe/H] $\leq$ 0.2, and the error of $T_{eff}$ $\leq$ 300\,K / log$g$ $\leq$ 0.3 / [Fe/H] $\leq$ 0.3. A certain proportion of small error stars can support this conclusion.

For the stellar evolutions of medium mass stars, the theory of a gap area in the H-R diagram is reflected at the red end of the lower panel of Fig. 1 and the upper panel of Fig. 4 around log$g$ = 3.2. In the upper panel of Fig. 4, the peak around log$g$ = 4.2 corresponds to the MS stars. The peak around log$g$ = 2.4 corresponds to the HB or RGB stars for stellar evolutions of small mass stars, the HB or RGB or AGB stars for stellar evolutions of medium mass stars. Statistics show that the metal abundances among A, F, G, and K type stars have a wide distribution. The metal abundance is an important parameter in stellar physics. In a fine research work, we can not simply replace the metal abundance of other stars with the metal abundance of the Sun. The grid values of metal abundance for different types of stars can be referred to Fig. 8.

The statistical results indicate that the rich metal abundances should be considered in the study of hot A type MS stars. Most A type stars have log$g$ from 3.5 to 4.5 according to the top panel of Fig. 5. There are 78,141 stars with $T_{eff}$ $\geq$ 7,500\,K and most of them are hot A type MS stars. We show the percentage histogram of the number of the 78,141 stars with the error of $T_{eff}$, log$g$, and [Fe/H] distribution in Fig. 14. There are total 66\%, 88\%, and 93\% of stars with the error of $T_{eff}$ $\leq$ 100\,K / log$g$ $\leq$ 0.1 / [Fe/H] $\leq$ 0.1, the error of $T_{eff}$ $\leq$ 200\,K / log$g$ $\leq$ 0.2 / [Fe/H] $\leq$ 0.2, and the error of $T_{eff}$ $\leq$ 300\,K / log$g$ $\leq$ 0.3 / [Fe/H] $\leq$ 0.3. A considerable proportion of small error stars can support this conclusion.

\section{Acknowledgment}

Guoshoujing Telescope (the Large Sky Area Multi-Object Fiber Spectroscopic Telescope LAMOST) is a National Major Scientific Project built by the Chinese Academy of Sciences. Funding for the project has been provided by the National Development and Reform Commission. LAMOST is operated and managed by the National Astronomical Observatories, Chinese Academy of Sciences. We acknowledge the support of NSFC through grant 11803004.

\label{lastpage}

\begin{thebibliography}{99}
\bibitem[\protect\citeauthoryear{Abazajian}{2003}]{b1} Abazajian K., Adelman-McCarthy J. K., Ag$\ddot{u}$eros M. A., et al., 2003, AJ, 126, 2081
\bibitem[\protect\citeauthoryear{Angulo}{1999}]{b1} Angulo C., Arnould M., Rayet M., et al., 1999, NuPhA, 656, 3
\bibitem[\protect\citeauthoryear{Astier}{2006}]{b1} Astier P., Guy J., Regnault N., et al., 2006, A\&A, 447, 31
\bibitem[\protect\citeauthoryear{Caughlan}{1988}]{b1} Caughlan G. R., Fowler W. A., 1988, ADNDT, 40, 283
\bibitem[\protect\citeauthoryear{Cassisi}{2007}]{b1} Cassisi S., Potekhin A. Y., Pietrinferni A., et al., 2007, ApJ, 661, 1094
\bibitem[\protect\citeauthoryear{Corsico}{2020}]{b1} C$\acute{o}$rsico A. H., 2020, FrASS, 7, 47
\bibitem[\protect\citeauthoryear{Cui}{2012}]{b1} Cui X. Q., Zhao Y. H., Chu Y. Q., et al., 2012, RAA, 12, 1197
\bibitem[\protect\citeauthoryear{Durrell}{1993}]{b1} Durrell P. R., Harris W. W., 1993, AJ, 105, 1420
\bibitem[\protect\citeauthoryear{Faulkner}{1966}]{b1} Faulkner J., 1966, ApJ, 144, 978
\bibitem[\protect\citeauthoryear{Ferguson}{2005}]{b1} Ferguson J. W., Alexander D. R., Allard F., et al., 2005, ApJ, 623, 585
\bibitem[\protect\citeauthoryear{Gianninas}{2011}]{b1} Gianninas A., Bergeron P., Ruiz M. T., 2011, ApJ, 743, 138
\bibitem[\protect\citeauthoryear{Gunn}{2006}]{b1} Gunn J. E., Siegmund W. A., Mannery E. J., et al., 2006, AJ, 131, 2332
\bibitem[\protect\citeauthoryear{Howell}{2014}]{b1} Howell S. B. Stobeck C., Haas M., et al., 2014, PASP, 126, 398
\bibitem[\protect\citeauthoryear{Iben}{1975}]{b1} Iben I., 1975, ApJ, 196, 525
\bibitem[\protect\citeauthoryear{Iglesias}{1993}]{b1} Iglesias C. A., Rogers F. J., 1993, ApJ, 412, 752
\bibitem[\protect\citeauthoryear{Iglesias}{1996}]{b1} Iglesias C. A., Rogers F. J., 1996, ApJ, 464, 943
\bibitem[\protect\citeauthoryear{Kulebi}{2011}]{b1} K$\ddot{u}$lebi B., Jordan S., Euchner F., et al., 2009, A\&A, 506, 1341
\bibitem[\protect\citeauthoryear{Luo}{2015}]{b1} Luo A. L., Zhao Y. H., Zhao G., et al., 2015, RAA, 15, 1095
\bibitem[\protect\citeauthoryear{Paxton}{2011}]{b1} Paxton B., Bildsten L., Dotter A., et al., 2011, ApJS, 192, 3
\bibitem[\protect\citeauthoryear{Potekhin}{2010}]{b1} Potekhin A. Y., Chabrier G., 2010, CoPP, 50, 82
\bibitem[\protect\citeauthoryear{Raymond}{2004}]{b1} Raymond S. N., Miknaitis G., Fraser J., et al., 2004, AJ, 127, 2978
\bibitem[\protect\citeauthoryear{Rogers}{2002}]{b1} Rogers F. J., Nayfonov A., 2002, ApJ, 576, 1064
\bibitem[\protect\citeauthoryear{Saumon}{1995}]{b1} Saumon D., Chabrier G., van Horn H. M., 1995, ApJS, 99, 713
\bibitem[\protect\citeauthoryear{Schaller}{1992}]{b1} Schaller G., Schaerer D., Meynet G., et al., 1992, A\&AS, 96, 269
\bibitem[\protect\citeauthoryear{Stoughton}{2002}]{b1} Stoughton C., Lupton R. H., Bernardi M., et al., 2002, AJ, 123, 485
\bibitem[\protect\citeauthoryear{Thompson}{2018}]{b1} Thompson S. E., Coughlin J. L., Hoffman K., et al., 2018, ApJS, 235, 38
\bibitem[\protect\citeauthoryear{Timmers}{2000}]{b1} Timmers F. X., Swesty F. D., 2000, ApJS, 126, 501
\bibitem[\protect\citeauthoryear{Tremblay}{2011}]{b1} Tremblay P. E., Bergeron P., Gianninas A., 2011, ApJ, 730, 128
\bibitem[\protect\citeauthoryear{Wu}{2014}]{b1} Wu Y., Du B., Luo A. L., et al., 2014, IAUS, 306, 340
\bibitem[\protect\citeauthoryear{Yakovlev}{1980}]{b1} Yakovlev D. G., Urpin V. A., 1980, SvA, 24, 303
\bibitem[\protect\citeauthoryear{Zhao}{2012}]{b1} Zhao G., Zhao Y. H., Chu Y. Q., et al., 2012, RAA, 12, 723

\end{thebibliography}
\end{document}